\documentclass[12pt]{article}
\usepackage{graphicx} 

\newcommand{\Det}{{\rm Det}}
\newcommand{\Tr}{{\rm Tr}}
\newcommand{\tr}{{\rm tr}}

\newcommand{\be}{\begin{equation}}
\newcommand{\ee}{\end{equation}}
\newcommand{\bea}{\begin{eqnarray}}
\newcommand{\eea}{\end{eqnarray}}
\newcommand{\ba}{\begin{array}{l}}
\newcommand{\ea}{\end{array}}

\newcommand{\re}[1]{(\ref{#1})}

\begin{document}

\begin{center}
{HEAVY-LIGHT QUARK SYSTEMS IN THE INSTANTON VACUUM}\footnote{Submitted to the Proceedings of the ISHEPP "Relativistic Nuclear Physics and Quantum Chromodynamics", 
JINR (Dubna), 4-9 October 2010.}
\\ 
Yousuf Musakhanov
\\
National University of Uzbekistan
\end{center}

\begin{abstract}

Assuming the gluon field is well approximated by instanton configurations we derive a partition function and calculate the specific correlators. 
Namely, the  heavy quark propagator and heavy quark-aniquark correlator with the account of the light quark determinant and QCD instanton vacuum properties.
With these knowledge we calculate the light quark contribution to the interaction between heavy quarks, which might be essential for the properties of a few heavy quarks systems like oniums and double-heavy baryons.

\end{abstract}

{\bf Introduction.}

The physics of the heavy mesons and baryons with open and hidden heavy
quarks is very reach and hot topic. Understanding the heavy-meson
physics is important for evaluation of the components of the $CKM$-matrix,
verification of the Standard Model and probing the physics beyond
it, as well as production of different exotic meson states. Currently
the experiments with $B$- and $D$-mesons are intensively studied
by Belle~\cite{Belle}, 
BaBar~\cite{BABAR}
and CDF collaborations, where unprecedented integrated luminocities
were achieved, as well as neutrino-production of open and hidden charm
in neutrino-hadron processes studied by K2K~\cite{K2K},
MiniBoone~\cite{MiniBooNE}, 
NuTeV~\cite{NuTeV}
and Minerva~\cite{Minerva} collaborations.

Theoretically, in pre-QCD era some success was achieved by the quantum-mechanical
models which use effective potentials to describe heavy hadrons and
their excitations (see e.g.~\cite{Eichten:1979ms} and references
therein). However, such description inevitably introduces undefined
phenomenological constants. The relation of these constants to QCD
parameters is quite obscure: due to interaction with gluons and virtual
light quark pairs all the constants contain nonperturbative dynamics.
The numerical values of these constants are determined from fits to
experimental data, which limits the predictive power of such models. 

An advanced version of the potential model is NRQCD~\cite{Bodwin:1994jh},
however in this model light quarks and their interactions with heavy
quarks via gluons is done in a phenomenological way. For this reason
it is limited to description of systems with two heavy quarks. Alternatively,
the heavy mesons are described in the Heavy Quark Effective Theory
(HQET) proposed in~\cite{Isgur:1989}, which treats
the heavy mesons using the pQCD methods but does not take into account
nonperturbative effects.

We propose to study the heavy quark physics in
the framework of the instanton vacuum model. This model was developed
in~\cite{Diakonov} and provided
a consistent description of the light mesons physics~\cite{Musakhanov}.

One of the most prominent advances of the instanton vacuum model 
is the correct description of the spontaneous breaking of the chiral
symmetry ($S\chi$SB), which is responsible for properties of most
hadrons and nuclei ~\cite{Leutwyler:2001hn}. The $S\chi$SB is due
to specific properties of QCD vacuum, which  is known to be one of
the most complicated objects due to perturbative as well as non-perturbative
fluctuations and is a very important object of investigations by methods
of Nonperturbative Quantum Chromo Dynamics (NQCD).  In the instanton
picture  $S\chi$SB is due to the delocalization of single-instanton
quark zero modes in the instanton medium. One of the advantages of
the  instanton vacuum is that it is characterized by only two parameters:
the average instanton size $\rho\sim0.3\,{\rm fm}$ and the average
inter-instanton distance $R\sim1\,{\rm fm}$. These essential numbers
were suggested in~\cite{Shuryak:1981ff} and were derived from $\Lambda_{\overline{{\rm MS}}}$
in ~\cite{Diakonov}. These values were recently confirmed
by lattice measurements \cite{lattice}.

In case of the heavy quarks, the instanton vacuum description was
discussed in~\cite{Diakonov:1989un,Chernyshev:1995gj}. For the heavy quarks 
even the charmed quark mass $m_{c}\sim1.5$~GeV is larger than the typical parameters
of the instanton media--the inverse instanton size $\rho^{-1}\approx600$~MeV
and the interinstanton distance $R^{-1}\approx200$~MeV and thus the quark mass determines the 
dynamics of the heavy quarks.

\textbf{Light quark determinant with the quark sources term.} 
 
 Instanton vacuum field is assumed as a superposition of $N_{+}$
instantons and $N_{-}$ antiinstantons: \begin{eqnarray}
A_{\mu}(x)=\sum_{I}^{N_{+}}A_{\mu}^{I}(\xi_{I},x)+\sum_{A}^{N_{-}}A_{\mu}^{A}(\xi_{A},x).\label{A}\end{eqnarray}
 Here $\xi=(\rho,z,U)$ are (anti)instanton collective coordinates--
size, position and color orientation (see reviews ~\cite{Diakonov,Schafer:1996wv}.
The main parameters of the model are the average inter-instanton distance
$R$ and the average instanton size $\rho$. The estimates of these
quantities are \begin{eqnarray}
 &  & \rho\simeq0.33\, fm,\, R\simeq1\,{\rm fm},\mbox{(phenomenological)}~~\mbox{\cite{Diakonov,Schafer:1996wv}},\label{classicalParameters}\\
 &  & \rho\simeq0.35\, fm,\, R\simeq0.95\,{\rm fm},\mbox{(variational)}~~\mbox{\cite{Diakonov}},\nonumber \\
 &  & \rho\simeq0.36\, fm,\, R\simeq0.89\,{\rm fm},~\mbox{(lattice)}~\mbox{\cite{lattice}}\nonumber \end{eqnarray}
 and have $\sim10-15\%$ uncertainty.


Our main assumption is the interpolation formula: 
\begin{eqnarray}\label{Si}
S_{i}=S_{0}+S_{0}\hat{p}\frac{|\Phi_{0i}><\Phi_{0i}|}{c_{i}}\hat{p}S_{0},\,\,\, S_{0}=\frac{1}{\hat{p}+im},\,\,\,
c_{i}=im<\Phi_{0i}|\hat{p}S_{0}|\Phi_{0i}>
\end{eqnarray}
 The advantage of this interpolation is shown by the projection of
$S_{i}$ to the zero-modes: \begin{eqnarray}
S_{i}|\Phi_{0i}>=\frac{1}{im}|\Phi_{0i}>,\,\,\,<\Phi_{0i}|S_{i}=<\Phi_{0i}|\frac{1}{im}\end{eqnarray}
 as it must be, while the similar projection of $S_{i}$ given by
~\cite{Diakonov} has a wrong component, negligible only in
the $m\rightarrow0$ limit.

Summation of the re-scattering series leads to the total quark propagator and making few further steps we get 
the fermionized representation of low-frequencies
light quark determinant in the presence of the quark sources, which is relevant for our problems, in the
form~\cite{Musakhanov}: 
${\Det}_{\rm low}\exp(-\xi^{+}S\xi)=$ 
\begin{eqnarray}
=\int\prod_{f}D\psi_{f}D\psi_{f}^{\dagger}\exp\int\sum_{f}\left(\psi_{f}^{\dagger}(\hat{p}\,+\, im_{f})\psi_{f}+\psi_{f}^{\dagger}\xi_{f}+\xi_{f}^{+}\psi_{f}\right)
\prod_{f} \prod_{\pm}^{N_{\pm}}V_{\pm,f}[\psi^{\dagger},\psi],
\label{part-func}
\end{eqnarray}
 where \begin{eqnarray}
V_{\pm,f}[\psi^{\dagger},\psi]=i\int d^{4}x\left(\psi_{f}^{\dagger}(x)\,\hat{p}\Phi_{\pm,0}(x;\zeta_{\pm})\right)\int d^{4}y\left(\Phi_{\pm,0}^{\dagger}(y;\zeta_{\pm})(\hat{p}\,\psi_{f}(y)\right),\label{V}\end{eqnarray}
  The averaging over collective coordinates $\xi_{\pm}$ is a rather simple procedure,
since the low density of the instanton medium ($\pi^{2}\left(\frac{\rho}{R}\right)^{4}\sim0.1$)
allows us to average over positions and orientations of the instantons
independently. 

Light quark partition function from \re{part-func} at $N_f=1$ and $N_\pm=N/2$ is exactly given by 
\bea
&&Z[\xi,\xi^+]=e^{-\xi^+\left(\hat p \,+\, i(m+M(p))\right)^{-1}\xi}
\exp\left[\Tr\ln\left(\hat p \,+\, i(m+M(p))\right)+N\ln\frac{N/2}{\lambda}-N\right]
\label{Z}
\\
&&N=\Tr\frac{iM(p)}{\hat p \,+\, i(m+M(p))},\,\,\, M(p)=\frac{\lambda}{N_c}(2\pi\rho F(p))^2.
\label{M}
\eea
Here the form-factor $F(p)$ is given by Fourier-transform of the zero-mode. The coupling  $\lambda$ and the dynamical quark mass $M(p)$ are  defined by the Eq. \re{M}.

 At $N_f >1$,  and in the saddle-point approximation (no meson loops contribution) $Z[\xi_f,\xi_f^+]$ has a similar form as the Eq. \re{Z}.

\textbf{Heavy quark propagator.} 

Define the heavy quark propagator as: 
\bea
&&S_H=\frac{1}{Z}\int D\psi D\psi^{\dagger} 
\left\{\prod_{\pm}^{N_{\pm}}\bar V_{\pm}[\psi^{\dagger} ,\psi ]\right\}\exp\int\left(\psi^{\dagger}(\hat p+im )\psi\right) w[\psi,\psi^\dagger]
\\ \nonumber
&&w[\psi,\psi^\dagger]
=\left\{\prod_{\pm}^{N_{\pm}}\bar V_{\pm}[\psi^{\dagger} ,\psi ]\right\}^{-1}\int D\zeta
\left\{\prod_{\pm}^{N_{\pm}}V_{\pm}[\psi^{\dagger} ,\psi ]\right\}\frac{1}{\theta^{-1}-\sum_i a_i},
\,\, w_\pm=\frac{1}{\theta^{-1}-a_\pm}, 
\\\nonumber
&&<t|\theta|t'>=\theta(t-t'),  <t|\theta^{-1}|t'>=-\frac{d}{dt}\delta(t-t'),
a_i(t)=iA_{i,\mu}(x(t))\frac{d}{dt}x_\mu(t)
\eea
 Accordingly \cite{Diakonov:1989un}
\bea
&&w^{-1}[\psi,\psi^\dagger]=\theta^{-1} + \frac{N}{2}\sum_\pm \frac{1}{\bar V_{\pm}[\psi^{\dagger} ,\psi ]}\int d\zeta_\pm\ V_{\pm}[\psi^{\dagger} ,\psi ]\left( \theta-a_\pm^{-1}\right)^{-1}+ O(N^2/V^2)
\nonumber\\
&&=\theta^{-1}-\frac{N}{2}\sum_{\pm}\bar V^{-1}_{\pm}[\psi^{\dagger} ,\psi ]\int d\zeta_\pm\ V_{\pm}[\psi^{\dagger} ,\psi ]\theta^{-1}(w_\pm-\theta)\theta^{-1}+ O(N^2/V^2)
\nonumber\\
&&=\theta^{-1} - \frac{N}{2}\sum_\pm \frac{1}{\bar V_{\pm}[\psi^{\dagger} ,\psi ]}\Delta_{H,\pm}[\psi^{\dagger},\psi ] + O(N^2/V^2).
\eea
and finally we get
\bea
\label{SH1}
&& S_H=\frac{1}{\theta^{-1} - \lambda\sum_\pm\Delta_{H,\pm}[\frac{\delta}{\delta\xi} ,\frac{\delta}{\delta\xi^+}] }
 \exp\left[-\xi^+\left(\hat p \,+\, i(m+M(p))\right)^{-1}\xi\right]_{|_{\xi=\xi^+=0}}
 \\
 &&\approx \frac{1}{\theta^{-1} - \lambda\sum_\pm\Delta_{H,\pm}[\frac{\delta}{\delta\xi} ,\frac{\delta}{\delta\xi^+}] 
 \exp\left[-\xi^+\left(\hat p \,+\, i(m+M(p))\right)^{-1}\xi\right]_{|_{\xi=\xi^+=0}}}
\label{SH2}
\\
&&S_H^{-1}\approx \theta^{-1}  
- i\tr\int \frac{d^4 k_1}{(2\pi)^4} \frac{\lambda(2\pi\rho )^2 F^2(k_1 ) }{N_c(\hat k_1 \,+\, i(m+M(k_1)))} 
 \frac{1}{2N_c}\sum_\pm\int d^4z_\pm  \tr_c\left(\theta^{-1}(w_\pm-\theta)\theta^{-1}\right)
 \nonumber\\
 &&=\theta^{-1} - \frac{N}{2VN_c}\sum_\pm\int d^4z_\pm  \tr_c\left(\theta^{-1}(w_\pm-\theta)\theta^{-1}\right)
\label{SH3}
\eea
The Eq. (\ref{SH3}) exactly coincide with the similar one from~\cite{Diakonov:1989un}.

Now re-write the Eq. (\ref{SH1}) introducing heavy quark fiels $Q,Q^\dagger$:
\bea
&& S_H=e^{\left[-\Tr\ln\left(\hat p \,+\, i(m+M(p))\right)\right]}\int D\psi D\psi^{\dagger}  D Q D Q^\dagger \,\,Q \, Q^\dagger\,\,\exp\left[\left(\psi^{\dagger}(\hat p +i(m+M(p)))\psi\right)+\right. 
\\\nonumber
&&\left.+  Q^\dagger\left(\theta^{-1} - \lambda\sum_\pm\Delta_{H,\pm}[\psi^{\dagger},\psi ] \right)Q-\Tr\ln\left(\theta^{-1} - \lambda\sum_\pm\Delta_{H,\pm}[\psi^{\dagger},\psi ]\right)\right],
\eea
where third term represent the (negligible) contribution of the heavy quark loops, while 
the second one is the heavy and light quarks interaction action, explicitly represented by
\bea
&& - \lambda\sum_\pm Q^\dagger\Delta_{H,\pm}[\psi^{\dagger},\psi ]Q= - i\lambda\sum_\pm\int d^4z_\pm \frac{d^4 k_1}{(2\pi)^4} \frac{d^4 k_2}{(2\pi)^4}  \exp(i(k_2-k_1)z_\pm) 
\\ \nonumber 
&&\times(2\pi\rho )^2 F(k_1 )F(k_2 )\left[ \frac{1}{N_c^2}\psi^+(k_1)\frac{1\pm\gamma_5}{2}\psi(k_2)Q^+ \tr_c\left(\theta^{-1}(w_\pm-\theta)\theta^{-1}\right)Q\right.
\\\nonumber 
&&\left.+\frac{1}{32(N_c^2-1)}\psi^+(k_1) (\gamma_\mu\gamma_\nu \frac{1\pm\gamma_5}{2})\lambda^i \psi(k_2)\tr(\tau^{\mp}_{\mu}\tau^{\pm}_{\nu}\lambda^j)
Q^+ \tr_c\left(\theta^{-1}(w_\pm-\theta)\theta^{-1}\lambda^j \right)\lambda^i Q\right]
\eea
{\bf Heavy quark anti-quark system.}

Define the correlator for this system as:
\bea
&&<T|C(L_1,L_2)|0>=\frac{1}{Z}\int D\psi D\psi^{\dagger} 
\left\{\prod_{\pm}^{N_{\pm}}\bar V_{\pm}[\psi^{\dagger} ,\psi ]\right\}\exp\int\left(\psi^{\dagger}(\hat p )\psi\right)W[\psi,\psi^\dagger]
\\ \nonumber
&&<T|W[\psi,\psi^\dagger]|0>
=\left\{\prod_{\pm}^{N_{\pm}}\bar V_{\pm}[\psi^{\dagger} ,\psi ]\right\}^{-1}\int D\zeta
\left\{\prod_{\pm}^{N_{\pm}}V_{\pm}[\psi^{\dagger} ,\psi ]\right\}
\\ \nonumber
&&<T|\left(\theta^{-1}-\sum_i a^{(1)}_i\right)^{-1}|0>
<0|\left(\theta^{-1}-\sum_i a^{(2)}_i\right)^{-1}|T>,
\eea
here the correlator is a Wilson loop along the rectangular contour  $L\times r$, where the sides $L_1,L_2$ are parallel to $x_4$ axes and 
separated by the distance $r$. The  $a^{(1)},a^{(2)}$ are the projections of the instantons onto the lines $L_1,L_2.$ 

Accordingly \cite{Pobylitsa:1989uq}
\bea
&&W^{-1}[\psi,\psi^\dagger]= w_1^{-1}[\psi,\psi^\dagger]\times w_2^{-1,T}[\psi,\psi^\dagger]
\\ \nonumber
&&-\frac{N}{2}\sum_\pm \bar V^{-1}_{\pm}[\psi^{\dagger} ,\psi ]\int d\zeta_\pm\ V_{\pm}[\psi^{\dagger} ,\psi ] \left(w_1[\psi,\psi^\dagger]-a^{(1)-1}_\pm\right)^{-1}\times\left(w_2[\psi,\psi^\dagger]-a^{(2)-1}_\pm\right)^{-1,T}
\\ \nonumber
&&= w_1^{-1}[\psi,\psi^\dagger]\times w_2^{-1,T}[\psi,\psi^\dagger]
\\ \nonumber
&&-\frac{N}{2}\sum_\pm \bar V^{-1}_{\pm}[\psi^{\dagger} ,\psi ]\int d\zeta_\pm\ V_{\pm}[\psi^{\dagger} ,\psi ] \left(\theta^{-1}\left(w^{(1)}_\pm-\theta\right)\theta^{-1}\right)\times\left(\theta^{-1}\left(w^{(2)}_\pm-\theta\right)\theta^{-1}\right)^{T}
\eea
where, superscript $T$ means the transposition, $\times$ -- tensor product and
\bea
&&w_1^{-1}[\psi,\psi^\dagger]=\theta^{-1}-\frac{N}{2}\sum_{\pm}\bar V^{-1}_{\pm}[\psi^{\dagger} ,\psi ]\int d\zeta_\pm V_{\pm}[\psi^{\dagger} ,\psi ]\theta^{-1}(w^{(1)}_\pm-\theta)\theta^{-1}+ O(N^2/V^2)
\nonumber\\
&&=\theta^{-1} - \frac{N}{2}\sum_\pm \frac{1}{\bar V_{\pm}[\psi^{\dagger} ,\psi ]}\Delta^{(1)}_{H,\pm}[\psi^{\dagger},\psi ] + O(N^2/V^2).
\eea
and similar for the $w_2^{-1}[\psi,\psi^\dagger].$

From previous calculations we see that the lowest orders on $\frac{N}{N_cV}$ in $C(L_1,L_2)$ are given by the integration over $\psi,\psi^\dagger$ of the  $W^{-1}[\psi,\psi^\dagger]. $ Then, we have the new interaction term between heavy quarks located on the lines $L_1$ and $L_2$ due to exchange of the light quarks between them.

Explicitly the integration of the first term in $W^{-1}[\psi,\psi^\dagger]$ over $\psi,\psi^\dagger$ leads to:
\bea
&&\frac{1}{Z}\int D\psi D\psi^{\dagger} 
\left\{\prod_{\pm}^{N_{\pm}}\bar V_{\pm}[\psi^{\dagger} ,\psi ]\right\}\exp\int\left(\psi^{\dagger}(\hat p )\psi\right)w_1^{-1}[\psi,\psi^\dagger]\times w_2^{-1,T}[\psi,\psi^\dagger]
\\\nonumber
&&=\left(\theta^{-1}-\lambda\sum_\pm\Delta^{(1)}_{H,\pm}[\frac{\delta}{\delta\xi} ,\frac{\delta}{\delta\xi^+}]\right) \times
\left(\theta^{-1}-\lambda\sum_\pm\Delta^{(2)}_{H,\pm}[\frac{\delta}{\delta\xi} ,\frac{\delta}{\delta\xi^+}]\right)^{T} {e^{-\xi^+\left(\hat p \,+\, i(m+M(p))\right)^{-1}\xi}}_{|_{\xi=\xi^+=0}}
\eea
Light quarks generated potential is given by
\bea
\nonumber
&&V_{lq}=\left(\lambda\sum_\pm\Delta^{(1)}_{H,\pm}[\frac{\delta}{\delta\xi_1} ,\frac{\delta}{\delta\xi_1^+}]\right) \times
\left(\lambda\sum_\pm\Delta^{(2)}_{H,\pm}[\frac{\delta}{\delta\xi_2} ,\frac{\delta}{\delta\xi_2^+}]\right)^{T} 
\\
&&\times\exp\left[-\xi_2^+\left(\hat p \,+\, i(m+M(p))\right)^{-1}\xi_1-\xi_1^+\left(\hat p \,+\, i(m+M(p))\right)^{-1}\xi_2\right]|_{\xi=\xi^+=0}
\eea
{\bf Conclusion.}

Approximating the gluon field by the instanton configurations
it was derived the low-frequency part of the light quark determinant in
the presence of quark sources.
 This one provided the way to calculate the heavy quark propagator with account of the light
quark determinant and QCD instanton vacuum properties and to derive instanon generated light-heavy quarks interaction terms. 
 With these knowledge it was calculated the light quark contribution to
the interaction between heavy quarks  providing more detailed investigation
of the few heavy quarks systems like oniums and double-heavy baryons and of the role 
of the spontaneous breaking of the chiral
symmetry (SBCS) in light quark sector for the heavy-light quarks systems.

\end{document}